\documentclass[a4paper,fleqn]{cas-sc}

\usepackage[numbers]{natbib}
\usepackage{gensymb}
\usepackage{subcaption}
\DeclareUnicodeCharacter{2212}{-}

\usepackage{enumitem}
\setlist{parsep=0pt,listparindent=\parindent}

\def\tsc#1{\csdef{#1}{\textsc{\lowercase{#1}}\xspace}}
\tsc{WGM}
\tsc{QE}
\tsc{EP}
\tsc{PMS}
\tsc{BEC}
\tsc{DE}


\begin{document}
\let\WriteBookmarks\relax
\def\floatpagepagefraction{1}
\def\textpagefraction{.001}

\shorttitle{Assessing the dark degeneracy through gas mass fraction data}

\shortauthors{Dinorah Barbosa et~al.}

\title [mode = title]{Assessing the dark degeneracy through gas mass fraction data}

%
\author[1]{Dinorah Barbosa}[orcid=0000-0002-7988-7454]
\cormark[1]
\cortext[cor1]{Corresponding author}
\ead{dinorahteixeira@on.br}

\author[2,3]{Rodrigo von Marttens}[orcid=0000-0003-3954-5756]
\ead{rodrigomarttens@ufba.br}
\author[4]{Javier Gonzalez}[orcid=0000-0002-5165-9783]
\ead{javiergonzalezs@academico.ufs.br}
\author[1]{Jailson Alcaniz}[orcid=0000-0003-2441-1413
]
\ead{alcaniz@on.br}

\affiliation[1]{organization={Observatório Nacional},
    city={Rio de Janeiro - RJ},
    postcode={20921-400}, 
    country={Brazil}}

\affiliation[2]{organization={Universidade Federal da Bahia},
    city={Salvador - BA},
    postcode={40170-110}, 
    country={Brazil}}

\affiliation[3]{organization={PPGCosmo, Universidade Federal do Espírito Santo},
    city={Vitória - ES},
    postcode={29075-910}, 
    country={Brazil}}

\affiliation[4]{organization={Universidade Federal de Sergipe},
    city={São Cristóvão - SE},
    postcode={49107-230}, 
    country={Brazil}}

\credit{Data curation, Writing - Original draft preparation}

\begin{abstract}
It is well-known that Einstein's equations constrain only the total energy-momentum tensor of the cosmic substratum, without specifying the characteristics of its individual constituents. Consequently, cosmological models featuring distinct decompositions within the dark sector, while sharing identical values for the sum of dark components' energy-momentum tensor, remain indistinguishable when assessed through observables based on distance measurements. Notably, it has been already demonstrated that cosmological models with dynamical descriptions of dark energy, characterized by a time-dependent equation of state (EoS), can always be mapped into a model featuring a decaying vacuum ($w=-1$) coupled with dark matter. We explore the possibility of breaking this degeneracy by using measurements of the gas mass fraction observed in massive and relaxed galaxy clusters. This data is particularly interesting for this purpose because it isolates the matter contribution, possibly allowing the degeneracy breaking. We study the particular case of the $w$CDM model with its interactive counterpart. We compare the results obtained from both descriptions with a non-parametric analysis obtained through Gaussian Process. Even though the degeneracy may be broken from the theoretical point of view, we find that current gas mass fraction data seems to be insufficient for a final conclusion about which approach is favored, even when combined with SNIa, BAO and CMB.
\end{abstract}


\begin{keywords}
cosmology \sep dark energy \sep dark matter \sep gas mass fraction 
\end{keywords}

\maketitle

\section{Introduction}

One of the main goals of modern cosmology is to understand the physics behind dark matter and dark energy, which together form the enigmatic dark sector. Within the standard $\Lambda$CDM model, whereas dark matter (DM) is considered cold, i.e., constituted by non-relativistic particles, dark energy (DE) is described by the cosmological constant $\Lambda$, which is equivalent to a perfect fluid with an equation of state (EoS) parameter $w = -1$. In this context, the most natural candidate for DE is a gravitating vacuum, however the discrepancy between the observed and expected value for the vacuum energy density is above $100$ orders of magnitude~\citep{weinberg1989cosmological,padmanabhan2003cosmological}. This outstanding difference is addressed as the ``cosmological constant problem'', and it is one of the most important open problems in Cosmology. Furthermore, the $\Lambda$CDM model also faces other problems, both on theoretical and observational scopes. In the theoretical sphere, we can highlight the problem of cosmic coincidence, which consists on the question of why dark matter and dark energy have only become comparable in recent times~\cite{Velten:2014nra}. On the other hand, in the observational side, current observational results indicate a considerable discrepancy with well-established CMB results~\cite{Planck:2018vyg}. The most relevant discrepancy is the $~5\sigma$ inconsistency between the local measurements from SH0ES~\cite{Riess:2021jrx} and the Planck results, the so-called $H_0$ tension~\cite{di2021realm}.

In response to these challenges, a plethora of alternative models has emerged in the literature, mainly focused on modifying the underlying physical mechanisms driving the recent cosmic acceleration, with the aim of resolving or alleviating the issues of the $\Lambda$CDM model. An important part of these cosmological models are defined by specific hypothesis on the dark components, e.g. the EoS of the dark energy component and/or a non-gravitational coupling between the components. However, due to the fact that the Einstein equations only constraint the sum of the energy-momentum tensors of their components, but do not impose any individual restrictions on each component, there exists a degeneracy in the sense that different models have identical geometrical properties, the so-called \textit{dark degeneracy}~\cite{Carneiro:2014uua}. For the particular case where dynamical and interacting dark energy models are considered, Ref.~\cite{von2020dark} showed that it is possible to find a correspondence between both approaches so that, for a dynamical model, we can determine its degenerated interacting counterpart. We can also employ these relations to explore the viability of dynamical models that mimic models with an interaction in the dark sector, as shown in~\cite{von2023one}. 

In this paper, we perform a test for breaking this degeneracy at the background level using gas mass fraction data from galaxy clusters, $f_{gas}(z)$. We benefit from the fact that this dataset offers measurements on $\Omega_{c}/\Omega_{m}$ rather than distances, which are function of $H(z)$. This dataset provides a test for distinguishing the values of cosmological parameters on both approaches, thus breaking the dark degeneracy. 
We perform a MCMC parameter estimation for $f_{gas}(z)$ alone, and its combination with SNIa, BAO and CMB.
For the purpose of determining which model is favored by the $f_{gas}(z)$ data, we compare our statistical analyses with a model-independent $\Omega_{c,0}$ estimate from the reconstruction of $\rho_{c}(z)$. For the dynamical approach, we consider a time-varying dark energy model where CDM evolves with $\rho_{cdm} \propto a^{-3}$ and, for the interactive model, a dark sector where decaying vacuum is coupled to dark matter. More precisely, we examine the case of flat $w$CDM and its correspondent interactive model, denoted as $\tilde{w}$CDM. 

This paper is divided as follows: In Sec.~\ref{sec:darkdegeneracy} we introduce the cosmological dark degeneracy while in Sec~\ref{sec:wcint} we detail the background and perturbation equations for the $\Tilde{w}$CDM. Sec~\ref{sec:methodology} describes the data used and analysis performed in the paper. In Sec.~\ref{sec:results} we report and discuss our main results. Finally, we conclude in Sec~\ref{sec:conclusions}. 

\section{Dark Degeneracy}
\label{sec:darkdegeneracy}

The Einstein's field equations establish a relation between the geometry of space-time and the constituents of the universe,
\begin{equation}
    \frac{G_{\mu\nu}}{\kappa} = \sum_{i} T_{\mu\nu}^i,
    \label{eq:einsteinfield}
\end{equation}
where $G_{\mu\nu}$ is the Einstein tensor, $\kappa= 8\pi G\slash c^4 $ is the universal gravitational constant, and $T_{\mu\nu}^i$ is the energy-momentum tensor of the $i-th$ component of the universe. $G_{\mu\nu}$ is determined for a specific gravitational theory, and the contributions of $T_{\mu\nu}$ for baryonic matter and radiation are known through their physical properties that we know from the laboratory. Thus, Eq.~\eqref{eq:einsteinfield} can always be rewritten in terms of the energy-momentum tensor for the dark sector, $T_{\mu\nu}^d = T_{\mu\nu}^{c} + T_{\mu\nu}^{x}$:
\begin{equation}
    T_{\mu\nu}^{d} =  \frac{G_{\mu\nu}}{\kappa}-T_{\mu\nu}^{b} - T_{\mu\nu}^{r}.
\end{equation}
Superscript $c$, $x$, $b$, and $r$ stand for dark matter, dark energy, baryons and radiation, respectively. Since we cannot directly measure the individual contributions of the dark sector components, cosmological models with the same value of $T_{\mu\nu}^{d}$ but different decompositions will be degenerated. This feature is known as the \textit{dark degeneracy}~\cite{Carneiro:2014uua,von2020dark,wasserman2002degeneracy,rubano2002quintessence,kunz2009degeneracy,aviles2011dark}. 

At the background level, the dark degeneracy translates to models having identical Hubble rate as a function of time. Therefore, observations that depend exclusively on this quantity (e.g., distance measurements) will yield identical results. It is noteworthy that in degenerate models, the sums of energy densities among dark components are identical, yet they evolve distinctly over time. In practical terms, this implies that degenerate models will share the same $H_0$, but will have different values for their dark matter and dark energy density parameters. However, as will be discussed in Sec.~\ref{subsec:degenerated}, there is always an algebraic relationship between the energy density solutions of the degenerate models. This degeneracy can also be maintained at the linear level, depending on the choice of dark energy sound speed and anisotropic shear. In this work, we fix the rest-frame dark energy sound speed as luminal and assume a spatially flat cosmology for both approaches.

\subsection{Degenerated models}
\label{subsec:degenerated}

Following Ref.~\cite{von2020dark}, we consider scenarios where a non-minimal coupling between dark energy and dark matter is allowed, while the other constituents, i.e., baryons and radiation remain unaffected by this relation and follow the standard evolution $\rho_{b}\propto a^{-3}$ and $\rho_{r}\propto a^{-4}$, respectively. Under these circumstances, even if the energy density of CDM and DE are not individually conserved due to energy exchange, the dark sector as whole is. Therefore, assuming a flat FLRW metric, a generalized fluid description of the dark sector has an equation of state:
\begin{equation}
    w_{d} = \frac{w_{x}}{1 + r(a)},
\end{equation}
where $r(a) \equiv \rho_{c}/\rho_{x}$ is the CDM/DE ratio. The dark fluid obeys the conservation equation
\begin{equation}
    \dot{\rho}_{d} + 3H[1 + w_{d}]\rho_{d} = 0\;,
\end{equation}
with $p_{d}= p_{x}$ and $\rho_{d}= \rho_{c} +\rho_{x}$. The general solution for $\rho_{d}$ can be written as
\begin{equation}
    \rho_{d} = \rho_{d,0}\,\mathrm{exp}\Big[-3\int\frac{1 + w_{d}(a')}{a'}da'\Big].
\end{equation}

One can find relations between two different dark sector decompositions by imposing that the left side of Eq.(~\ref{eq:einsteinfield}) is the same for both approaches. We center on interactive and non-interactive cases for the dark sector. We consider a standard dynamical approach for dark energy where it is characterized by a time-dependent equation of state $w_x(a)$. As for the interactive approach, we study couplings between a cosmological constant ($\Tilde{w}_x = -1$) and dark matter. We adopt an \textit{ansatz} for the interaction source function that depends exclusively on the energy densities of the constituents of the dark sector as discussed in~\cite{vonMarttens:2018iav},
\begin{equation}
    Q = 3HR(\Tilde{\rho}_{c},\Tilde{\rho}_{x})\ ,
\label{eq:q_int}
\end{equation} 
with $R(\Tilde{\rho}_{c},\Tilde{\rho}_{x})$ a function of $\Tilde{\rho}_{c}$ and $\Tilde{\rho}_{x}$, the energy densities for CDM and DE in the interacting model, respectively.

In either case, we describe the dark sector as single fluid that is independent of the other components of the universe, as in Sec.~\ref{subsec:degenerated}. Hereafter, we indicate cosmological constituents of the dynamical model with a bar superscript, and do the same for the interacting case with a tilde. In the event where the quantity is the same for both models, we simply omit this notation. 

Forcing the conditions of the dark degeneracy at background level implies that the degenerated models must have the same $w_{d}(a)$. This can be expressed as the following:
\begin{equation}
  \Bar{w}_{x}\frac{\Bar{\rho}_{x}}{\Bar{\rho}_{x} + \Bar{\rho}_{c}} = - \frac{\Tilde{\rho}_{x}}{\Tilde{\rho}_{x} + \Tilde{\rho}_{c}}.
  \label{eq:degen_background}
\end{equation}
On the other hand, the total energy density must also be equal $\rho_{d} = \Bar{\rho}_{c} + \Bar{\rho}_{x} = \Tilde{\rho}_{c} + \Tilde{\rho}_{x}$. Thus, Eq. (~\ref{eq:degen_background}) reduces to
\begin{equation}
    \Tilde{\rho}_{x} = -\Bar{w}_{x}\Bar{\rho}_x\,,
    \label{eq:rhox_int}
\end{equation}
and combining the last two equations we also find an expression for $\Tilde{\rho}_{c}$:
\begin{equation}
    \Tilde{\rho}_{c} = \Bar{\rho}_{c} + (1 + \Bar{w}_{x})\Bar{\rho}_{x}.
    \label{eq:rhoc_int}
\end{equation}

Notice that the above expressions explicitly demonstrate how we can move from one scenario to another. We are particularly interested in finding the interacting counterpart for a given dynamical model. Nevertheless, the inverse way is also possible~\cite{von2023one}, that is, for a given interaction in the form of Eq. (\ref{eq:q_int}), we are able to find its dynamical equivalent. 
%
Furthermore, from Eqs. (\ref{eq:rhox_int}) and (\ref{eq:rhoc_int}):
\begin{equation}
    \Tilde{\Omega}_{x,0} = -\Bar{w}_0\Bar{\Omega}_{x,0},\qquad{\rm and}\qquad \Tilde{\Omega}_{c,0} = \Bar{\Omega}_{c,0} + (1 + \Bar{w}_0)\Bar{\Omega}_{x,0},
    \label{eq:rho_tilde}
\end{equation}
which illustrates how, even with a degeneracy at the background level, the values for the density parameters of each dark species for the two cases will not be identical, i.e. $\Bar{\Omega}_{x,0}\neq\Tilde{\Omega}_{x,0}$; $\Bar{\Omega}_{c,0}\neq\Tilde{\Omega}_{c,0}$. This is expected, since the physics described for the dark sector is different between the models and, therefore, the only case that they will be equal is for a vanishing interaction or equivalently for a non-evolving dark energy fluid ($\Bar{w}(z)=-1$). Additionally, since we have set $\Tilde{w}_x = -1$ for the interactive case, the only $w_0$ that appears in the solutions for both the dynamical and interactive case is $\Bar{w}_0$. Thus, we choose to drop the bar notation for this parameter. It is noteworthy that, while $w_0$ is the usual EoS parameter for the dark energy in the dynamical model, for the interactive model it plays a role of an interaction parameter in the sense that $w_0 = -1$ means a vanishing interaction. Likewise, the bigger the departure of $w_0$ from $-1$ for the interactive case, the stronger the interaction in the dark sector for this approach.

Similarly, the conditions for the dark degeneracy in the linear level can also be obtained from Einstein and conservation equations. In this case, the degeneracy will depend on the sound speed of the dark energy and the anisotropic stress.
In this work, we follow the procedure elaborated in Ref.~\cite{von2020dark} for the linear level, where the speed of sound of dark energy in its comoving reference frame is fixed at 1 in both approaches. Although this choice breaks the degeneracy at the linear level, it was shown in Ref.~\cite{von2020dark} that from a practical point of view, the degenerate models remain indistinguishable using CMB data.

\section{Specific case: $w$CDM \textit{vs.} $\Tilde{w}$CDM}
\label{sec:wcint}
Considering what was discussed in the previous section, we study the specific case of a flat interacting model that mimics flat $w$CDM, designated as $\Tilde{w}$CDM. For the $w$CDM model, the background solutions for the energy densities of the dark components are given by
\begin{equation}
    \Bar{\rho}_{c} = \frac{3H_0^2}{8\pi G}\Bar{\Omega}_{c,0}a^{-3}\qquad{\rm and}\qquad \Bar{\rho}_{x} = \frac{3H_0^2}{8\pi G}\Bar{\Omega}_{x,0}a^{-3(1 + w_0)}\,.
\end{equation}
On the other hand, for the $\Tilde{w}$CDM model we have,
\begin{equation}
    \Tilde{\rho}_{c} = \frac{3H_0^2}{8\pi G}[\bar{\Omega}_{c,0} + \bar{\Omega}_{x,0}(1 + w_0)a^{-3w_0}]a^{-3}\qquad{\rm and}\qquad \Tilde{\rho}_x = -\frac{3H_0^2}{8\pi G}\bar{\Omega}_{x,0}w_0a^{-3(1 + w_0)}\,.
    \label{eq:back_wcdmint2}
\end{equation}

For the perturbations, we assume that the anisotropic stress and pressure perturbation for the CDM are zero. As previously mentioned, the comoving speed of sound of DE is fixed in both approaches by setting $c_{s}^2 = \Bar{c}_{s}^2 = \Tilde{c}_{s}^2 = 1$. In this context, the general perturbative equations are,
\begin{equation}
\begin{split}
\delta_i^{'} & + 3\mathcal{H}(c_{s(i)}^{2} - w_i)\delta_{i} + (1 + w_i)(\theta_{i} + 3\Phi^{'}) \\
 & + 3\mathcal{H}[3\mathcal{H}(1 + w_{i})(c_{s(i)}^2 - w_i) + w_i^{'}]\frac{\theta_{i}}{k^2} = \frac{aQ_{i}}{\rho_{i}}\Big[\Psi - \delta_i + 3\mathcal{H}(c_{s(i)}^{2} - w_i)\frac{\theta_i}{k^2} + \frac{\delta Q_i}{Q_i}\Big],
\end{split}
\label{eq:per_delta}
\end{equation}
and 
\begin{equation}
\theta_i^{'} + \mathrm{H}(1 - 3c_{s(i)}^2)\theta_i - \frac{c_{s(i)}^{2}}{1 + w_i}k^2\delta_{i} + k^2\sigma_{i} - k^2\Psi = \frac{aQ_i}{\rho_i(1 + w_i)}\Big[\theta - (1 + c_{s(i)}^2)\theta_i - \frac{k^2\mathcal{F}_i}{Q_i}\Big].
\label{eq:per_theta}
\end{equation}
The equations above are written in terms of the density contrast $\delta$ and comoving sound speed $c_{s(i)}^2$. Here, the interaction source function is (see e.g. Eq. (65) of Ref.\cite{von2020dark})
\begin{equation}
    Q = \frac{3H\Tilde{\rho}_{c}\Tilde{\rho}_{x}}{\Tilde{\rho}_{c} + \Tilde{\rho}_{x}}\Big[-1 -\frac{\Bar{r}_{0}w_{0}a^{3w_0}}{1 + w_0 + \Bar{r}_{0}a^{3w_0}}\Big],
    \label{eq:q_wcdm}
\end{equation}
while the interacting momentum transfer $\mathcal{F}$ is defined to obtain a geodesic CDM~\cite{Borges:2017jvi}, as well as in~\cite{von2023one}.

\section{Methodology}
\label{sec:methodology}

To discuss a possible breaking of dark degeneracy at the background level, we use datasets of complementary samples to perform our observational analysis of the dynamical and interacting approaches to dark energy. 

\subsection{Observational data}
\label{subsec:data}

The cosmological observational data employed in this work are:
\begin{enumerate}

    \item{\textbf{fgas: }}In massive galaxy clusters, the hot gas in the intracluster medium (ICM) is the main source of X-ray emission from the cluster. This allows us to estimate the gas mass fraction $f_{gas} = M_{gas}/M_{tot}$. Hydrodynamical simulations suggest that $f_{gas}(z)$ remains roughly constant in time and has low cluster-to-cluster scatter~\cite{eke1998evolution, crain2007baryon, kay2004cosmological, battaglia2013cluster, planelles2013baryon}. Thus, it can be used to obtain cosmological parameters~\cite{sasaki1996new}. The gas mass fraction catalog used in this work is taken from Mantz et al. \cite{mantz2014cosmology} (hereafter M14). It contains 40 measurements of gas mass fraction calculated in a spherical shell of  $0.8 < r/r^{ref}_{2500} <1.2$ from the center of the cluster:
\begin{equation}
    f_{gas}^{ref}\Bigg(0.8<\frac{r}{r_{2500}^{ref}} < 1.2 ;z \Bigg) = K(z)A(z)\Upsilon(z)\Bigg(\frac{\Omega_{b,0}}{\Omega_{m,0}}\Bigg)\Bigg[\frac{d^{ref}_A(z)}{d_A(z)}\Bigg]^{3/2},
    \label{eq:fgas}
\end{equation}
%
with $\Omega_{m} = \Omega_{b} + \Omega_{c}$. $K(z)$ is the X-ray to weak lensing mass relation, given by 
\begin{equation}
    K(z) = K_0(1 + K_1z),
\end{equation}
$\Upsilon(z)$ the gas depletion
\begin{equation}
    \Upsilon(z) = \Upsilon_0(1 + \Upsilon_1z),
\end{equation}
and $A(z)$ the angular correction relative to the cosmological model of reference
\begin{equation}
    A(z) \approx \Bigg(\frac{H(z)d(z)}{H^{ref}(z)d^{ref}(z)}\Bigg)^{\eta}.
\end{equation}
    For M14, the fiducial cosmology assumed is the flat $\Lambda$CDM with $\Omega_{m,0}=0.30$ and $H_0=70$ km/s/Mpc. Additionally, a weak lensing mass calibration of the X-ray measurements is often performed in order to enhance the $f_{gas}(z)$ data. This procedure directly affects the final value of $K_0$, while $K_1$ is usually unconstrained by the data. It is also noted that $K_0$ has a small dependence on the cosmological model assumed~\cite{applegate2016cosmology}. 

    \item{\textbf{SN Ia: }}The physical mechanisms that describe type Ia supernova allows us to treat them standarizable candles, therefore obtaining cosmological distance measurements. For the supernovae Ia (SNIa) data, we use the Pantheon compilation~\cite{scolnic2018complete}. This sample contains 1048 SNIa in the redshift range $0.01 \leq z \leq 2.3$.
    The distance modulus is given by $\mu = m_{B} + M$, where $m_{B}$ is the apparent magnitude in the $B$ band, and $M$ is the absolute magnitude. It relates to the luminosity distance $d_L$ through
    \begin{equation}
    \mu = 5\ log\Big(\frac{d_L}{10pc}\Big),
    \end{equation}
    where 
\begin{equation}
    d_L = (1+z)\int_{0}^{z}\frac{dz'}{H(z')},
\end{equation}
    for a flat universe. 

    \item{\textbf{BAO: }}The baryon acoustic oscillations (BAO) of the photon-baryon fluid in the early universe left an imprint in the three-dimensional distribution of galaxies, known as the BAO feature. The BAO relates to a characteristic sound-horizon at the baryon drag epoch, $r_d$:
\begin{equation}
    r_d = \frac{c}{\sqrt{3}}\int_{0}^{a_d}\frac{da}{a^2H(a)\sqrt{1 + \frac{3\Omega_{b,0}}{\Omega_{\gamma,0}}a}},
\end{equation}
    where $a_d$ is the scale factor at the drag epoch, $\Omega_{\gamma,0}$ is the current density parameter for photons. Moreover, information about the BAO peak can be obtained from the compressed quantities: (i) the angular feature $D_M/r_d$, (ii) the radial feature $D_H/r_d$, with $D_M= (1 + z)D_A$, $D_H= c/H$ the characteristic distance for the respective features. On the other hand, on scales where the peculiar velocities of galaxies are negligible, we can obtain information about the redshift space distortions (RSD) that are sensitive to the structure formation of the universe, such as the quantity $f\sigma_8$.
    
    For the joint BAO/RSD we use galaxy/quasar clustering data from the following combination of SDSS DR7, BOSS DR12 and eBOSS DR16: SDSS DR7 main galaxy sample (MGS)~\cite{Ross:2014qpa}, BOSS DR 12 Luminous Red Galaxy (LRG) catalog~\cite{BOSS:2016wmc}, eBOSS DR 16 LGR sample, eBOSS DR16 quasar catalog (QSO), the Lya auto- and cross-correlation power spectrum from the eBOSS DR 16 sample~\cite{eBOSS:2020yzd}. We do not use the full power-spectrum but the aforementioned compressed information from the BAO as well as $f\sigma_8$ from RSD. We also consider these datasets to be independent, but a covariance matrix is considered for different measurements inside each catalog. More details about this data combination, including a table with the specific quantities, measurements, mean redshifts and errors can be found in~\cite{andrade2021test}.

    \item{\textbf{CMB: } The cosmic microwave background (CMB) is a relic of the time when photons and baryons decoupled and the universe became transparent. This radiation has since cooled down and currently has a blackbody spectrum temperature of about $T\simeq 2.73$ K~\cite{fixsen2009temperature}. The CMB is highly isotropic across the sky\footnote{Considering the removal of the CMB dipole, located at $(l,b) \approx (264\degree,48\degree)$~\cite{aghanim2013planck}.}. The anisotropies in the temperature map of the CMB is of the order $10^{-5}$. These inhomogeneties maps allow us to construct auto- and cross-correlation power spectrum of temperature, polarization and lensing for different angular scales. Information contained in the power spectrum, such as angular scale and relative height of the acoustic peaks are directly related to cosmological parameters, which makes CMB a powerful probe in constraining cosmological models~\cite{aghanim2020planck}. Additionally, as the photons of the CMB travel across space they are deflected by weak gravitational lensing and therefore must undergo a lensing reconstruction to correct these effects~\cite{lewis2006weak, planck2018lensing}.
    
    In this work, we use the \texttt{PlanckTTTEEE\_lensing} from the Planck 2018 archive~\cite{aghanim2020planck}, which has measurements for the temperature and E-mode polarization auto- and cross-correlation power spectrum, as well as the lensing reconstruction. For high multipoles, we adopted the \texttt{Plik} likelihood. It covers multipoles from $l\geq30$ up to $\sim2500$ for TT and $30\leq l \leq 1996$ for TE,EE. We refer the reader to~\cite{planck2020likelihood} for more details about this likelihood code.}

\end{enumerate}

\subsection{Statistical Analysis}
\label{subsec:statistical}
We use modified versions of the Boltzmann solver CLASS~\cite{lesgourgues2011cosmic} to compute the background evolution for our models and perform a Monte Carlo analysis with the \texttt{MontePython} code~\cite{Audren:2012wb}. We created a likelihood code \textbf{in \texttt{MontePython}} to use the $f_{gas}(z)$ data from M14. Our likelihood does not perform the calibration between X-ray data and weak lensing. Instead, following the simplified analysis reported in~\cite{mantz2022cosmological}, we use an external prior on $K_{0}$ from the Weighting the Giants project (WtG)~\cite{applegate2016cosmology}. 
For simplicity, we chose not to include the intrinsic scatter at $f_{gas}(r_{2500})$, which is reportedly low ($\sim7\%$)~\cite{mantz2014cosmology}, and the mass dependency $M_{2500}$ of the gas mass fraction. 

To evaluate the effect of adding gas mass fraction data to break the degeneracy for the background, we performed three distinct analyses for each model, separating the background data from CMB: $f_{gas}(z)$, $f_{gas}(z)+\mathrm{SNIa}+\mathrm{BAO}$ and $f_{gas}(z)+\mathrm{CMB}$. The priors chosen for each parameter in our $f_{gas}(z)$ analysis are described in Table~\ref{tab:priors_fgas}. The external priors used for $100\,\omega_{b,0}$ and $h$ are from Cooke et al.(2014)~\cite{cooke2014precision} and Riess 2021a~\cite{riess2021cosmic}, hereafter R21. When using $f_{gas}(z)$ data in combination with other datasets, we do not add the external priors for $h$, $100\,\omega_{b,0}$, but maintain $K_0$ and $\eta$. We used the standard cosmological parameters and their respective priors from this Planck 2018 dataset for the CMB data. We also used \texttt{GetDist}~\cite{lewis2019getdist} to analyze and plot the chains. 
 
\begin{table}[h!]
    \centering
    \begin{tabular}{c c c}
         \hline\\
         & Parameter & Prior \\
         \hline 
         \rule{0pt}{3ex} 
         & Cosmological & \\
         
         $\star$ & $100\,\omega_{b,0}$ & $\mathcal{N}[2.202;0.046]$ \\
         $\star$ & $h$ & $\mathcal{N}[0.732;0.013]$\\
         & $\Omega_{c,0}$ & $\mathcal{U}[0;1]$\\
         & $w_0$ & -\\
         \rule{0pt}{3ex} 
         & Astrophysical & \\
         $\star$ & $K_0$ & $\mathcal{N}[0.96;0.09]$\\
         & $K_1$ & $\mathcal{U}[-0.05;0.05]$\\
         & $\Upsilon_0$ & $\mathcal{U}[0.763;0.932]$\\
         & $\Upsilon_1$ & $\mathcal{U}[-0.05;0.05]$\\
         $\star$ & $\eta$ & $\mathcal{N}[0.442;0.035]$ \\
         \hline
    \end{tabular}
    \caption{Priors used for our MCMC data analysis for $f_{gas}(z)$. Gaussian priors are noted with $\mathcal{N}$[Mean; Standard deviation], while uniform priors are notes as $\mathcal{U}$[Min; Max]. Where a star symbol ($\star$) appears alongside the parameter name, it means the prior was external. For the EoS parameter $w_0$, for both the dynamical and interacting models, we used a wide prior.}
    \label{tab:priors_fgas}
\end{table}

\section{Results}
\label{sec:results}

\subsection{Statistical analysis}
\label{subsec:results_fgas}

\begin{figure}[h!]
    \centering
    \includegraphics[scale = 0.60]{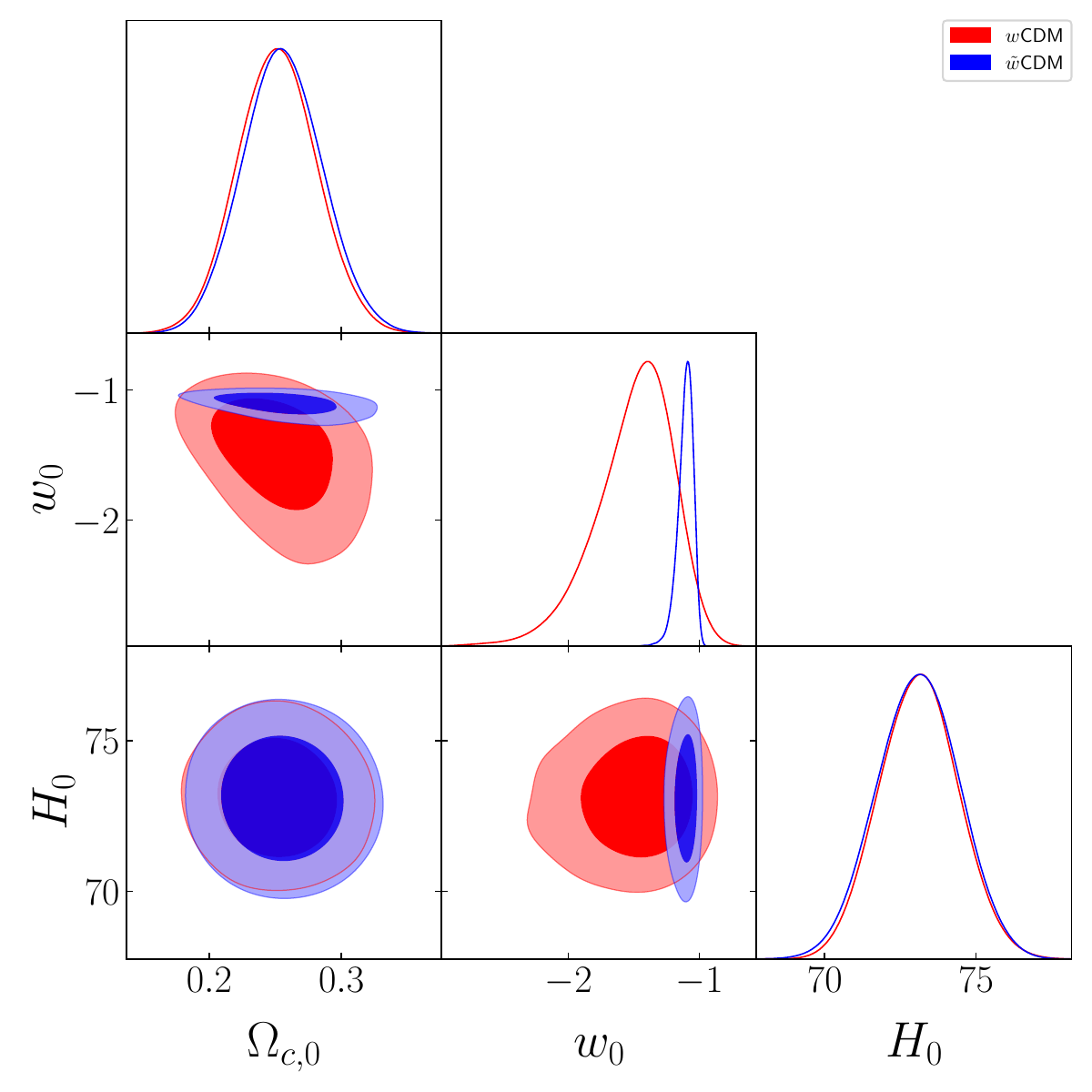}
        \caption{Triangle plot for $f_{gas}(z)$ data. $w$CDM is shown in red, whereas $\Tilde{w}$CDM is shown in blue. While $\Omega_{c,0}$ is essentially the same in both approaches, $w_0$ do not agree at $1\sigma$. A phantom dark energy for $w$CDM is in agreement with~\cite{mantz2014cosmology}~\cite{mantz2022cosmological}. For $\Tilde{w}$CDM, the tight constraints for $w_0$ are due to the fact that this parameter appears directly in the expression for $\Tilde{\Omega}_{c,0}$, as opposed to the dynamical case. On the other hand, the posterior distribution for $H_0$ is essentially a reproduction of the prior, as $f_{gas}(z)$ data imposes weak constraints on this quantity. We find the statistical differences between the two models to be insufficient to break the dark degeneracy in this analysis.}
    
    \label{fig:fgas}
\end{figure}

\begin{figure}[h]
    \centering
    \includegraphics[scale = 0.6]{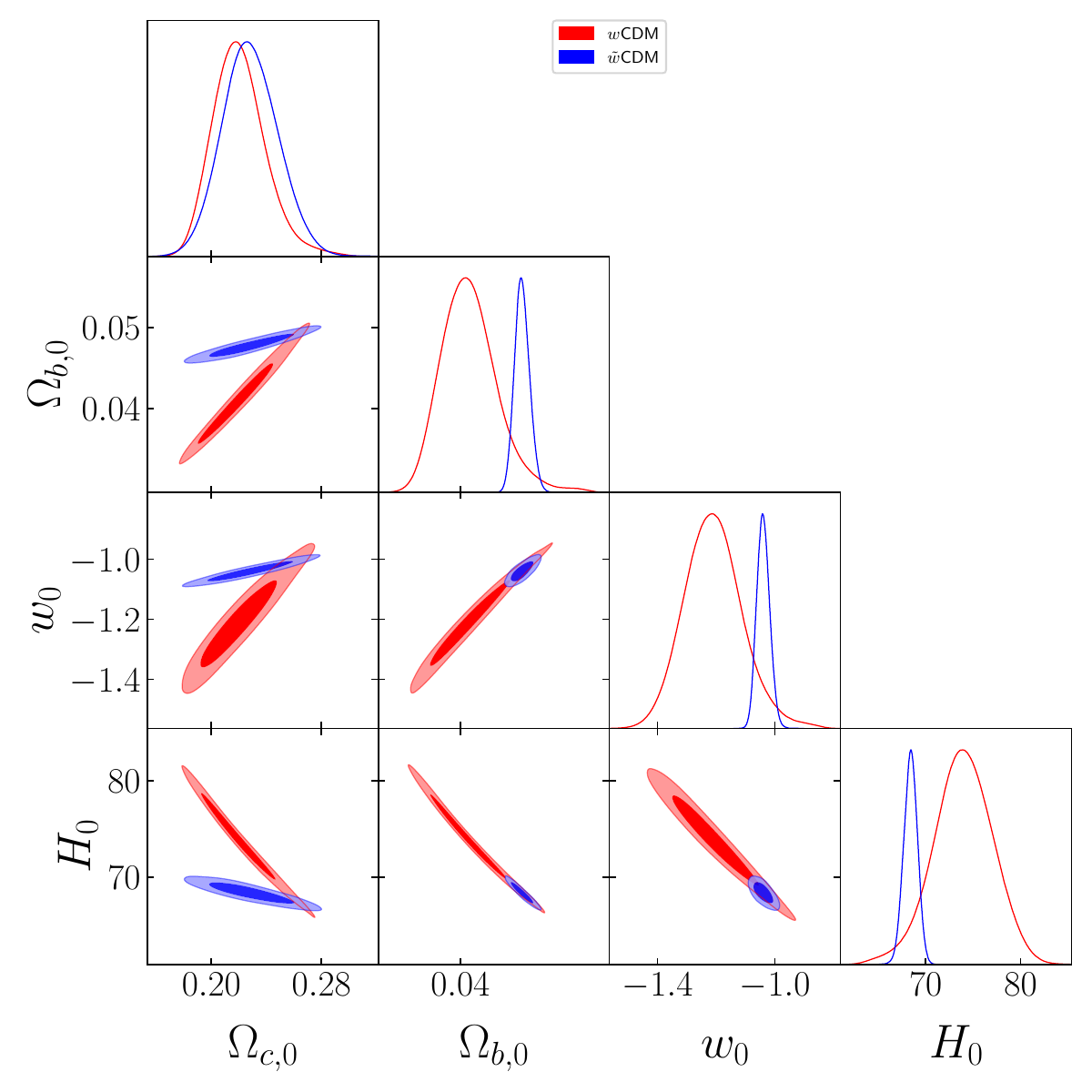}
    \caption{Triangle plot for $f_{gas}(z)+\mathrm{CMB}$ data. $w$CDM is shown in red, whereas $\tilde{w}$CDM is shown in blue. Differences between the two approaches are expected for this analysis, given our choice to break the degeneracy at linear scales. Once again, the phantom $w_0$ for $w$CDM is consistent with the findings for both $f_{gas}(z)$ and CMB~\cite{aghanim2020planck}.}
    \label{fig:fgas_cmb}
\end{figure}

We present the results for our statistical analysis in Figs.~\ref{fig:fgas},~\ref{fig:fgas_bao} and~\ref{fig:fgas_cmb}. The mean and $1\sigma$ values for the $\Omega_{c,0}$ and $w_0$ can be seen in Table~\ref{tab:bestfit}. For the $f_{gas}(z)$, $\Omega_{c,0}$ does not  vary considerably in both analysis, however $w_0$ is significantly different in $w$CDM to $\Tilde{w}$CDM, although both models are in agreement with $\Lambda$CDM at $2\sigma$ confidence level. 

A phantom dark energy at $1\sigma$ is expected for $w$CDM, as $f_{gas}(z)$ data shows preference for lower values of $w_0$ for this particular model~\cite{mantz2014cosmology,mantz2022cosmological}. Additionally, one can notice that the error bars for $\Tilde{w}$CDM in this analysis are smaller than for the dynamical case. This can be explained by the fact that in the dynamical model the parameter $w_0$ only appears inside $H(z)$ in $d_{A}(z)$ and $A(z)$ and therefore constraints on this parameter are weak. In contrast, for the interactive model, $w_0$ appears in the expressions for $\Tilde{\Omega}_{c}$, which results in  tighter constraints from this dataset. Furthermore, since we find $w_0$ to be close to $w_0=-1$, it suggests a weak interaction for this model. We also note from Fig. ~\ref{fig:fgas} that $H_0$ is similar for both models, but in the same way that $w_0$ is poorly constrained for $w$CDM by $f_{gas}(z)$, $H_0$ is also. Therefore, the $H_0$ results essentially reproduce the adopted prior for this parameter. 
Given the small differences between the statistical analyses for $f_{gas}(z)$, we find that the degeneracy was not broken at a background level for this dataset.

\subsection{Combination with other probes}
\label{subsec:results_other}

\begin{figure}[h]
    \centering
    \includegraphics[scale = 0.6]{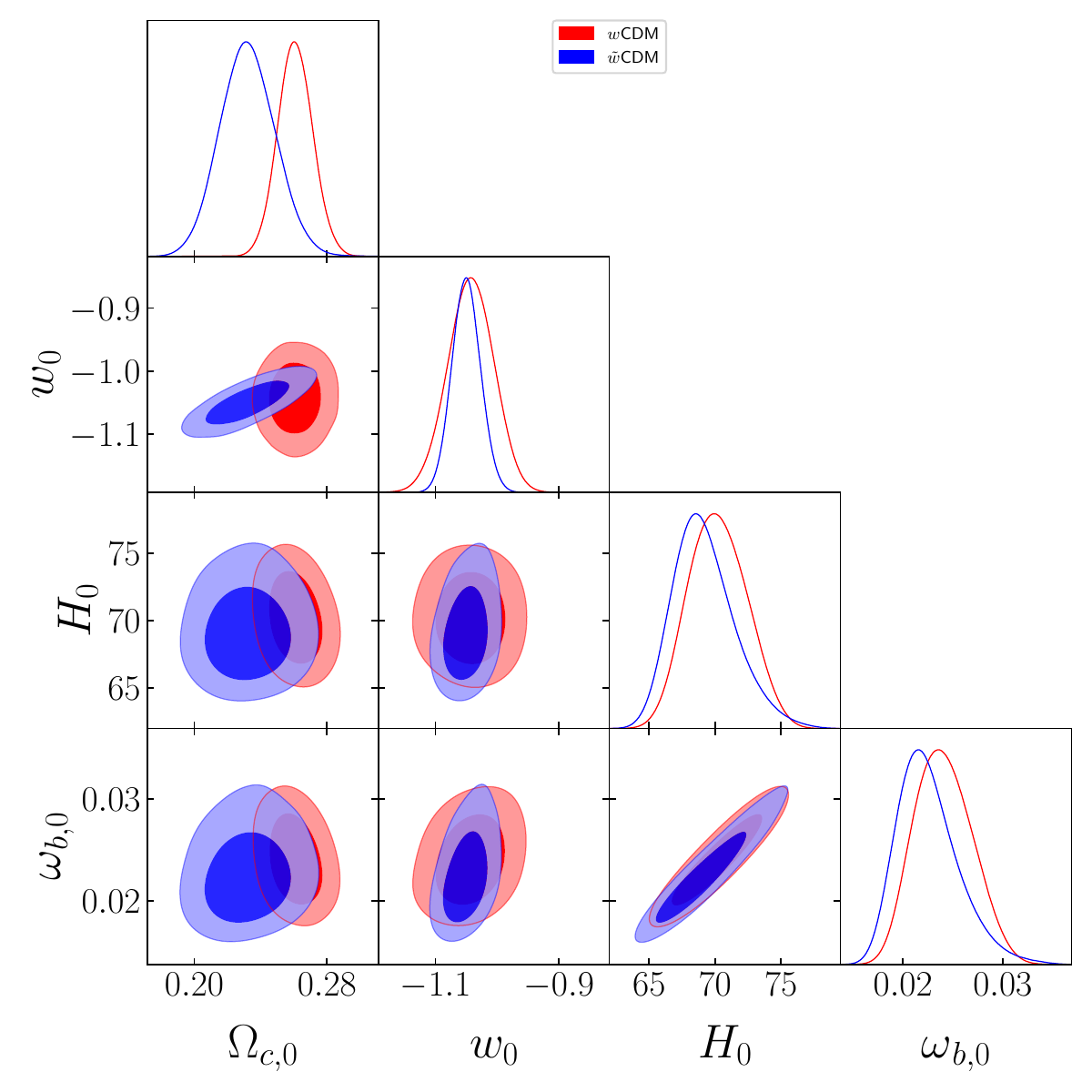}
    \caption{Triangle plot for $f_{gas}(z)+\mathrm{SNIa}+\mathrm{BAO}$ data. $w$CDM is shown in red, whereas $\tilde{w}$CDM is shown in blue. Now, $w_0$ substantially overlaps, while differences in $\Omega_{c,0}$ for the two approaches become more apparent. Although the errorbars for $w_0$ in $\Tilde{w}$CDM remain tight, the errorbars for the other parameters increase due to error propagation, in regards to $f_{gas}(z)$ only. Just as in the previous analysis, we find no substantial evidence for breaking the dark degeneracy with $f_{gas}+\mathrm{SNIa}+\mathrm{BAO}$.}
    \label{fig:fgas_bao}
\end{figure}

When incorporating $f_{gas}(z)$ to $\mathrm{SNIa}+\mathrm{BAO}$ the differences between $w_0$ shift notably in regards to the previous case. While this data combination provides more distinct values for $\Omega_{c,0}$, $w_0$ substantially overlaps. We also performed this same analysis for $f_{gas}(z)+\mathrm{SNIa}$ and $f_{gas}(z)+\mathrm{BAO}$ and found similar results. Furthermore, the constrains for $w_0$ remain smaller for $\Tilde{w}$CDM, but the error propagations for this model now play a bigger role constraining the other parameters, which can be visualized by the enlargement of the error bars for $\Tilde{w}$CDM in Fig.~\ref{fig:fgas_bao}. 

Lastly, for the $f_{gas}(z)+\mathrm{CMB}$, we find comparable values for $\Omega_{c,0}$, but $w_0$ diverges. These differences are expected, as the degeneracy is broken on linear level as per our choice for the comoving sound speed for dark energy. Moreover, we again find a phantom value for $w_0$ at more than $1\sigma$ for $w$CDM. We associate this result to the dataset of the analysis, noticing that, just as in the case for $f_{gas}(z)$ alone, CMB results find a phantom value for $w_0=-1.57^{+0.50}_{-0.40}$ (Table 4 in Ref.~\cite{aghanim2020planck}) when using the same data from Planck. Since both datasets show this preference for this model, the net effect also results in a lower $w_0$. For the $\Tilde{w}$CDM, we find that for either $f_{gas}+\mathrm{SNIa}+\mathrm{BAO}$ or $f_{gas}(z)+\mathrm{CMB}$, $w_0$ is consistent with a vanishing interaction ($w_0 = -1$) at $2\sigma$.
Furthermore, similarly to the analyses with only gas mass fraction, we find that combinations of $f_{gas}(z)$ with supernovae, BAO and CMB probes do no have an impact in breaking the dark degeneracy.

\renewcommand{\arraystretch}{1.25}
\begin{table}[h]
    \centering
    \begin{tabular}{c c c c }
    \hline
    \hline
    & Analysis & $\Omega_{c,0}$ & $w_0$ \\
    \hline
    \multirow{3}{4em}{$w$CDM}  & $f_{gas}(z)$ &  $0.252\pm 0.030$ & $-1.50^{+0.34}_{-0.22}$ \\
    &  $f_{gas}(z)+\mathrm{SNIa}+\mathrm{BAO}$ & $0.261\pm 0.011$  & $-1.043\pm 0.038 $ \\
    & $f_{gas}(z)+\mathrm{CMB}$ & $0.221^{+0.016}_{-0.021}$  & $-1.209^{+0.088}_{-0.10}$ \\
    \hline 
    \multirow{3}{4em}{$\Tilde{w}$CDM}  & $f_{gas}(z)$ & $0.256\pm 0.030$ & $-1.106^{+0.068}_{-0.042}$ \\
    &  $f_{gas}(z)+\mathrm{SNIa}+\mathrm{BAO}$ & $0.232\pm 0.017$ & $-1.050\pm 0.024$  \\
    & $f_{gas}(z)+\mathrm{CMB}$ & $0.228\pm 0.020$ & $-1.040^{+0.020}_{-0.023}$ \\
    \hline
    \end{tabular}
    \caption{Mean and $1\sigma$ values found for the analysis performed for $w$CDM and $\Tilde{w}\mathrm{CDM}$. When analyzing $f_{gas}(z)$ only, the models do not agree for the EoS parameter $w_0$ at $1\sigma$ C.L.. For $f_{gas}(z)+\mathrm{CMB}$, the variation in $w_0$ between the two approaches is expected, as we opted to break the degeneracy at linear scales.}
    \label{tab:bestfit}

\end{table}

\subsection{Model-independent determination of CDM energy density}
\label{subsec:gp_reconstruction}

\begin{figure}[h!]
\centering
\begin{subfigure}[t]{0.4\textwidth}
    \raisebox{1.4em}{\includegraphics[height = 5cm, width = \textwidth]{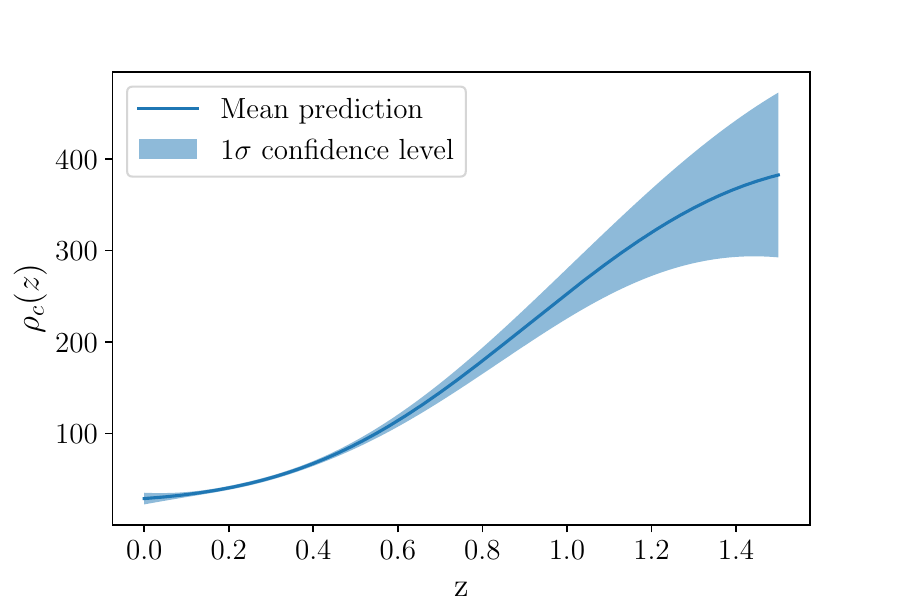}}
    \centering
    \caption{}
    \label{fig:M92}
\end{subfigure}\hspace{2em}
\begin{subfigure}[t]{0.4\textwidth}
    \includegraphics[height = 5cm, width = \textwidth]{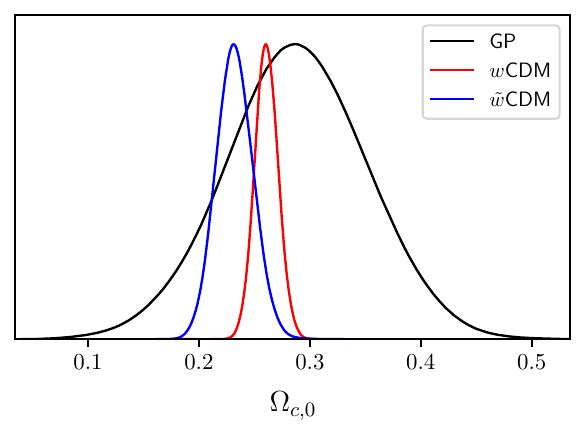}
    \centering
    \caption{}
    \label{fig:omega_M92}
\end{subfigure}
\caption{Results for GP using Matérn($9/2$). \textbf{(a):} values of the reconstructed $\rho_{c}$. Central values are denoted with a line, while the $1\sigma$ confidence level is represented by the shaded region. \textbf{(b):} result for $\Omega_{c,0}$ from M$92$ (in black), where $H_0=73\pm1.3\,\mathrm{km/s/Mpc}$ was employed. The red and blue curves are the results for $w$CDM and $\Tilde{w}$CDM, respectively for the data combination $f_{gas}(z)+\mathrm{SNIa}+\mathrm{BAO}$.
\label{fig:gp}
}
\end{figure}

For the purpose of assessing which model is favored by $f_{gas}(z)$ data, we compare our previous results with a model independent value of $\Omega_{c,0}$, derived from a Gaussian process (GP) reconstruction of $\rho_{c}(z)$. We rewrite Eq. (\ref{eq:fgas}) as
\begin{equation}
    \rho_{c}(z) = \rho_{b}(z)\Big[\frac{K(z)A(z)\Upsilon(z)}{f_{gas}^{ref}(z)}\Bigg(\frac{d_A^{ref}(z)}{d_A(z)}\Bigg)^{3/2} - 1\Big].
\end{equation}
We use the same 
priors displayed in Tab. \ref{tab:priors_fgas} and the M14 data to perform  the GP reconstruction. We reconstruct $\rho_c(z)$ using five kernels and chose the one that maximized the GP marginal likelihood, Matérn$(9/2)$ (M92). For this case, we found $\rho_{c,0}=29.1\pm6.4\cdot10^{-31}\mathrm{g}\slash \mathrm{cm}^{3}$, which translates into $\Omega_{c,0} = 0.29\pm0.06$, with $H_0=(73.2\pm1.3)\,\mathrm{km/s/Mpc}$ (R21). Table~\ref{tab:results_gp_all} shows the results for all kernels tested, although the results do not change significantly. An image of this reconstruction is shown in Fig. ~\ref{fig:M92}. A similar figure containing all the kernels used in this work is presented in Fig.~\ref{fig:gp_all}. 

We calculate the discrepancy $D = {\lvert X - Y \rvert}/{\sqrt{\sigma_X^2 + \sigma_Y^2}}$, where $X$ and $Y$ are the mean values of a given quantity and $\sigma_X$ and $\sigma_Y$ their respective errors, between our models' predictions for $\Omega_{c,0}$ and the model-independent value from GP. We find $0.53\sigma$ for $w$CDM and $0.47\sigma$ for $\tilde{w}$CDM in regards to the $f_{gas}(z)$ results.
As presented in Tab.~\ref{tab:bestfit} and~\ref{fig:fgas}, the differences between the two approaches are small for this parameter, and exhibit no statistical relevance.
Tab.~\ref{tab:tension_gp} shows the discrepancies for the other samples. 
For the background $f_{gas}(z)+\mathrm{SNIa}+\mathrm{BAO}$ we have the biggest difference between the discrepancies, which is expected, as this data combination provides the most contrasting values of $\Omega_{c,0}$ for both models.
On the other hand, the highest discrepancies are found for $f_{gas}(z)+\mathrm{CMB}$.
Overall, the discrepancies for the models are insignificant to indicate a preference for any of the cosmologies tested. This is attributed to the large errors associated to the current $f_{gas}(z)$ data and the similarities for the statistical analyses between $w$CDM and $\Tilde{w}$CDM. 

\begin{table}[h!]
    \centering
    \begin{tabular}{c c c }
    \hline
    \hline
    & Analysis & Discrepancy($\sigma$) \\
    \hline
    \multirow{3}{4em}{$w$CDM}  & $f_{gas}(z)$ &  $0.53$ \\
    &  $f_{gas}(z)+\mathrm{SNIa}+\mathrm{BAO}$ &  $0.43$ \\
    & $f_{gas}(z)+\mathrm{CMB}$ & $1.04$ \\
    \hline 
    \multirow{3}{4em}{$\Tilde{w}$CDM}  & $f_{gas}(z)$ & $0.47$ \\
    & $f_{gas}(z)+\mathrm{SNIa}+\mathrm{BAO}$ & $0.86$ \\
    & $f_{gas}(z)+\mathrm{CMB}$ & $0.91$  \\
    \hline
    \end{tabular}
    \caption{Discrepancies for $w$CDM and $\Tilde{w}$CDM in comparison with the $\Omega_{c,0}$ from the GP reconstruction, using $H_0=73\pm1.3\,\mathrm{km/s/Mpc}$. The biggest difference between models is found for $f_{gas}(z)+\mathrm{SNIa}+\mathrm{BAO}$, while the highest discrepancies are found for $f_{gas}(z)+\mathrm{CMB}$. In general, we consider the discrepancies too small for a substantial preference for either $w$CDM or $\Tilde{w}$CDM.}
    \label{tab:tension_gp}

\end{table}

\section{Conclusions}
\label{sec:conclusions}

The dark degeneracy that arises from Einstein field equation makes it impossible for models with the same total value of energy-momentum tensor to be completely distinguished by certain types of data. This degeneracy can, however, be avoided at perturbative linear scales with the right choice of parameters. Ref.~\cite{von2020dark} developed connections between quantities of degenerated models with dynamical and interacting scenarios, where in the latter, the dark sector is constituted by dark matter interacting with a cosmological constant.

In this work, we explored the possibility of breaking this degeneracy in the background by employing $f_{gas}(z)$ data, which offers direct measurements $\Omega_{c}$. We studied one of the parameterizations discussed in Ref.~\cite{von2020dark}, the ($w$CDM, $\Tilde{w}$CDM) pair.  We performed a parameter estimation for different datasets and found that the degeneracy is not broken from the current $f_{gas}(z)$ data alone or combined with other probes. We compare these results with a model-independent value for $\Omega_{c,0}$ obtained through Gaussian Process by calculating the discrepancies for $\Omega_{c,0}$ in each analysis. Our results show no statistically significant preference for either of the evaluated cosmologies. We also find $\Tilde{w}$CDM to be consistent with a vanishing interaction at $\simeq2\sigma$ for all data combination employed.

Overall, we expect future $f_{gas}(z)$ measurements to be able to significantly lower the error bars of the data, hopefully making it a more powerful probe for breaking dark degeneracy when combined with other experiments.  Alternatively, if another type of background data independent of $H(z)$ only is combined, we expect the combination with $f_{gas}(z)$ also to give us more constraining results for the dark degeneracy.

\section*{Acknowledgments}
DB acknowledges financial support from the Coordena\c{c}\~ao de Aperfei\c{c}oamento de Pessoal de N\'{\i}vel Superior - CAPES. RvM is suported by Funda\c{c}\~ao de Amparo \`a Pesquisa do Estado da Bahia (FAPESB) grant TO APP0039/2023. JSA is supported by CNPq grant No. 307683/2022-2 and Funda\c{c}\~ao de Amparo \`a Pesquisa do Estado do Rio de Janeiro (FAPERJ) grant No. 259610 (2021). We also acknowledge the use of CLASS and MontePython codes. This work was developed thanks to the use of the National Observatory Data Center (CPDON).

\bibliographystyle{model1-num-names}

\newpage

\newpage

\appendix
\label{appendix}
\section{Additional triangle plots and GP for all kernels}
\label{app:fullchains}

\begin{figure}[h!]
    \centering
    \includegraphics[scale = 0.7]{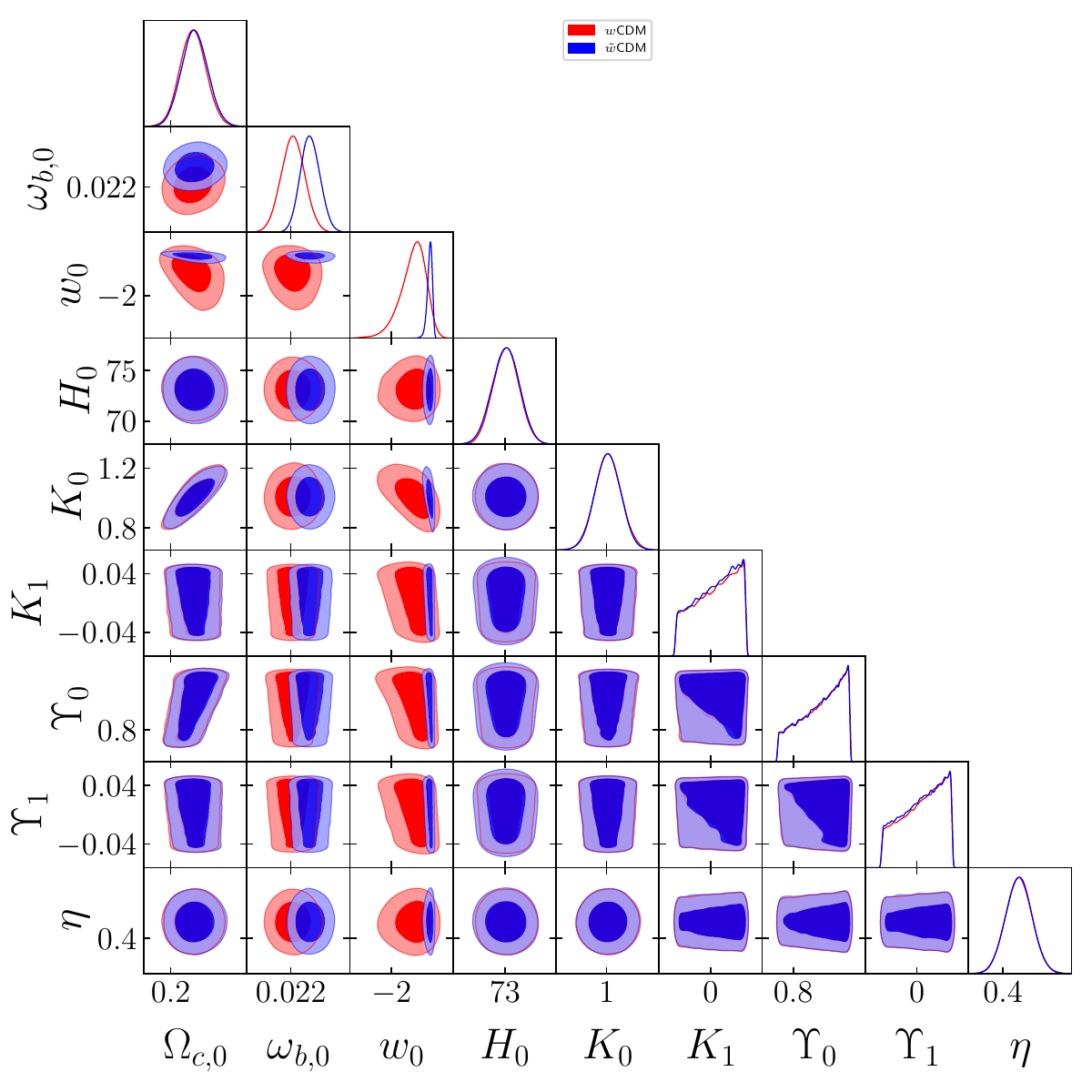}
    \caption{Full triangle plot for $f_{gas}(z)$. $w$CDM is shown in red, whereas $\tilde{w}$CDM is shown in blue. }
    \label{fig:fgas_triangle}
\end{figure}

\begin{figure}[h!]
    \centering
    \includegraphics[scale = 0.8]{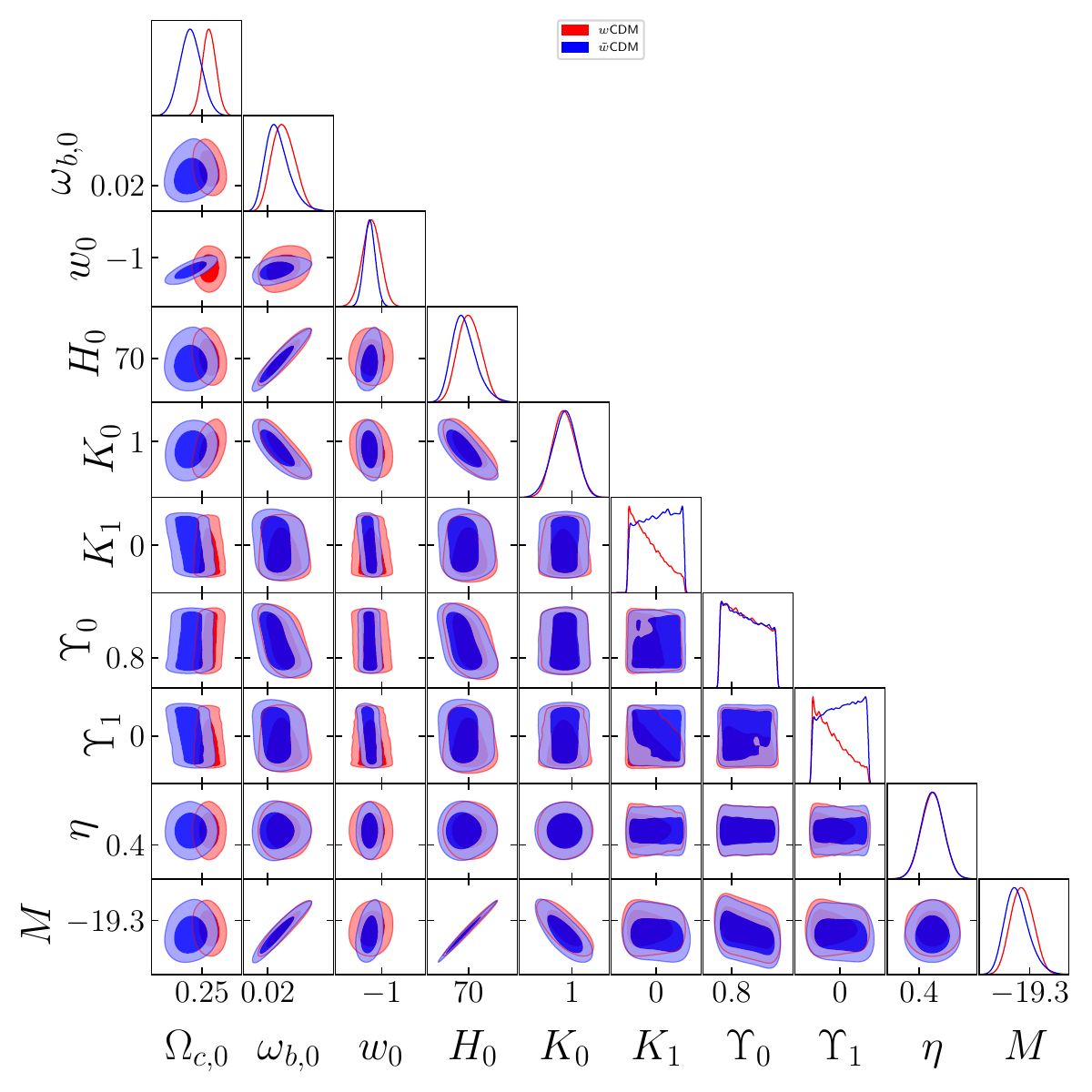}
    \caption{Full triangle plot for $f_{gas}(z)$+ BAO + SNIa data. $w$CDM is shown in red, whereas $\tilde{w}$CDM is shown in blue.}
    \label{fig:fgas_bao_sn_triangle}
\end{figure}

\begin{figure}[h!]
    \centering
    \includegraphics[scale = 0.6]{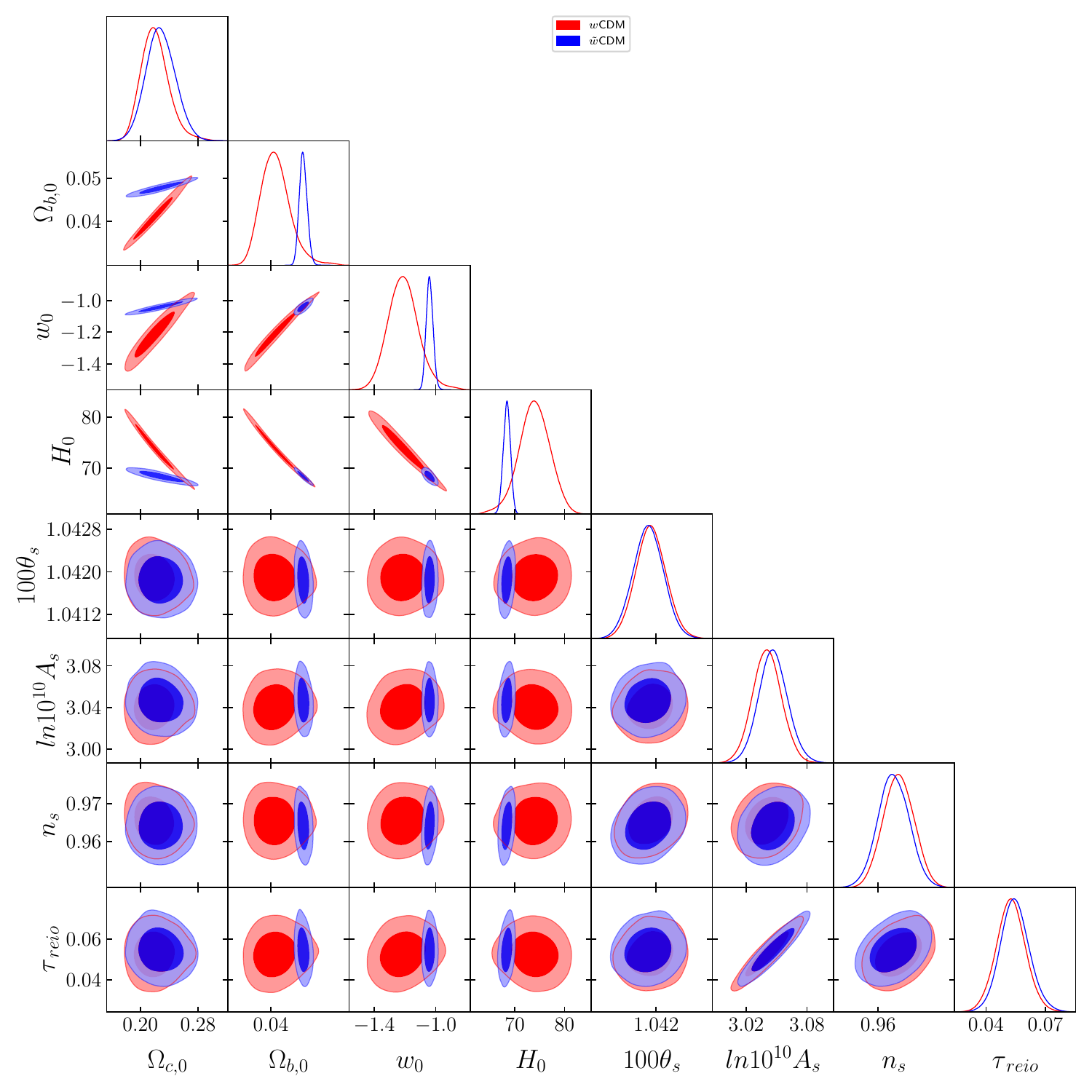}
    \caption{Triangle plot for $f_{gas}(z)$ + CMB data with nuisance parameters from CMB. $w$CDM is shown in red, whereas $\tilde{w}$CDM is shown in blue.}
    \label{fig:fgas_cmb_triangle}
\end{figure}

\newpage

\begin{figure}[h!]
    \centering
    \includegraphics[scale = 0.6]{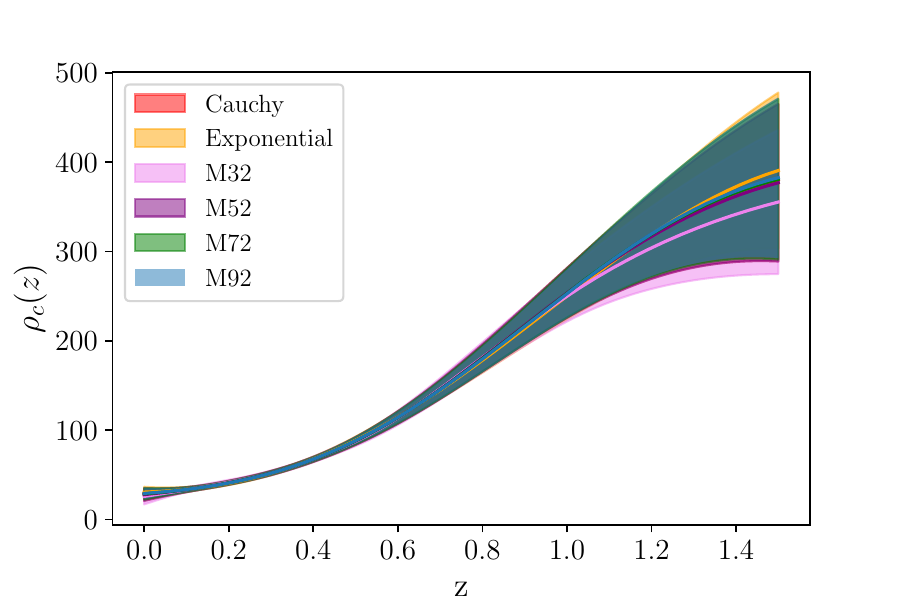}
    \caption{Gaussian process results with all kernels. $\mathrm{M}92$ stands for the kernel Matérn$(9/2)$, $\mathrm{M}72$ for Matérn$(7/2)$ and so on. As in Fig.~\ref{fig:M92}, the best-fit values are denoted with a line, and the $1\sigma$ confidence level is represented by the shaded region.}
    \label{fig:gp_all}
\end{figure}

\begin{table}[h!]
    \centering
    \begin{tabular}{c c c c c}
    \hline \\[-1.5ex]
    Kernel &  $\rho_{c,0}[10^{-31}\frac{g}{cm^{3}}]$ & $\omega_{c,0}$ & $\Omega_{c,0}$ & Loglike \\ [1.5ex]
    \hline \\[-1.5ex]
     M92 & $29.1 \pm 6.4$ & $0.14 \pm 0.03$ & $0.29\pm 0.06$ & $-161.1678$\\
     M72 & $28.6 \pm 6.6$ & $0.14 \pm 0.03$ & $0.28 \pm 0.07$ & $-161.2062$\\
     M52 & $27.7 \pm 7.0$ & $0.14 \pm 0.04$ & $0.28 \pm 0.07$ & $-161.3676$\\
     M32 & $25.7 \pm 8.9$ & $0.13 \pm 0.05$ & $0.26 \pm 0.09$ & $-162.2346$\\
     Exponential  & $30.6 \pm 6$ & $0.15 \pm 0.03$ & $0.30 \pm 0.06$ & $-161.1727$ \\
     Cauchy    & $29.1 \pm 6.4$ & $0.14 \pm 0.03$ & $0.29\pm 0.06$ & $-161.1766$ \\
    \hline 
    \end{tabular}
    \caption{Results for GP for different kernels. $\mathrm{M}92$ stands for the kernel Matérn$(9/2)$, $\mathrm{M}72$ for Matérn$(7/2)$ and so on. For $\Omega_{c,0}$, we used $H_0=(73.2 \pm 1.3)\,\mathrm{km/s/Mpc}$}.
    \label{tab:results_gp_all}
\end{table}

\end{document}